# Gold-diamond Nanocomposites Efficiently Generate Hydrated Electrons upon Absorption of Visible Light.

*Silvia Orlanducci[1], Giuseppe Ammirati[2,3], Alessandro Bellucci[4], Daniele Catone[2], Lionel C. Gontard[5], Faustino Martelli[6], Roberto Matassa[7], Alessandra Paladini[8], Francesco Toschi[2], Stefano Turchini[2], Patrick O'Keeffe[8,\**]*

[1]Dipartimento di Scienze e Tecnologie Chimiche, Università di Roma "Tor Vergata", Via della Ricerca Scientifica, 1, 00133 Rome, Italy.

[2]Istituto di Struttura della Materia-CNR (ISM-CNR), EuroFEL Support Laboratory (EFSL), 100 Via del Fosso del Cavaliere, 00133 Rome, Italy.

[3]CHOSE (Centre for Hybrid and Organic Solar Energy), Department of Electronic Engineering, University of Rome Tor Vergata, 00133 Rome, Italy.

[4]Istituto di Struttura della Materia-CNR (ISM-CNR), DiaTHEMA Lab, 00015 Monterotondo Scalo, Italy.

[5]Department of Condensed Matter Physics, and IMEYMAT, University of Cádiz, 11510 Puerto Real, Spain.

[6]Istitute of Microelectronics and Microsystems-CNR (IMM-CNR), 100 Via del Fosso del Cavaliere, 00133 Rome, Italy.

[7]Department of Anatomical, Histological, Forensic and Orthopaedic Sciences, Section of Human Anatomy, Sapienza University of Rome, Via A. Borelli 50, 00161 Rome, Italy.

[8]Istituto di Struttura della Materia-CNR (ISM-CNR), EuroFEL Support Laboratory (EFSL), 00015 Monterotondo Scalo, Italy.

*Corresponding author: patrick.okeeffe@ism.cnr.it

Table of Contents (TOC) Graphic

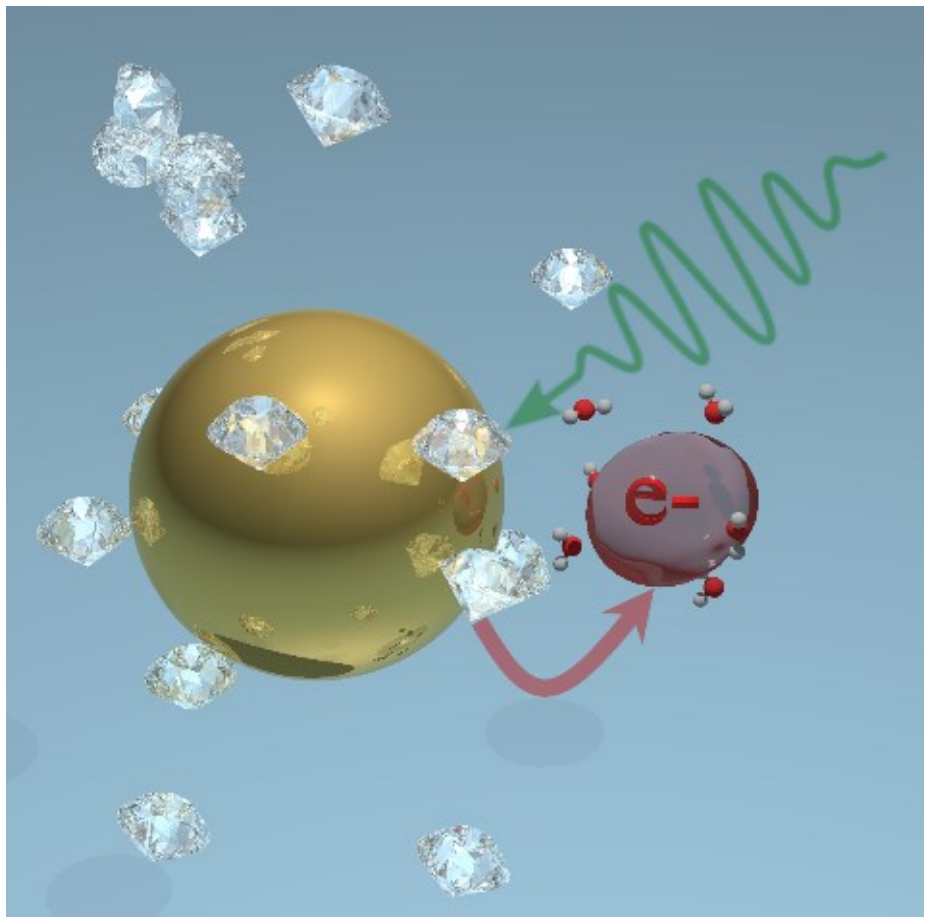

**ABSTRACT**

An efficient source of hydrated electrons generated by visible light has the potential to have a major impact on solar homogeneous catalysis. Diamond has potentially a high capability of emitting hydrated electrons, but only using ultraviolet light (λ<225 nm). In this work, we demonstrate the efficient absorption of visible light by nanocomposites consisting of detonation nanodiamonds and gold nanoparticles (AuNP@DNDs), which subsequently emit electrons into the aqueous environment in which they are suspended. This has been done by exciting the AuNP@DND with visible laser light and monitoring the appearance and intensity of the transient absorption of hydrated electrons centered at around 720 nm. We suggest that this mechanism is made possible by the plasmonic enhancement of visible absorption by $sp^2$-hybridized islands on the DND surface. Optimization of this process could lead to important breakthroughs in solar photocatalysis of energy-intensive reactions such as $N_2$ and $CO_2$ reduction as well as providing a non-toxic source of hydrated electrons for applications in wastewater management and nanomedicine.

## 1. INTRODUCTION

The production of solvated electrons, and more specifically, hydrated electrons has been extensively studied[1,2] since they were first optically identified by Hart et al. in the 1960s.[3] This is largely due to the fact that they have one of the highest reducing potentials of any chemical species.[2,4] This means that the hydrated electrons can promote many energy-demanding processes such as $CO_2$ reduction,[3] targeted chemical transformations,[5] and advanced reductive processes for applications in wastewater treatment.[6,7] Hydrated electrons have even been shown to promote fixation of $N_2$ through a multistep process.[8,9]

However, the conditions required to generate hydrated electrons are usually very energy intensive and/or toxic. Highly energetic methods include ionizing radiation (e.g. X-rays),[10] pulse radiolysis,[1] plasma discharge,[11] or multiphoton ionization using intense laser pulses.[12] Other possible sources of hydrated electrons include photoionization of the I⁻ ions in KI or NaI salt solutions which can produce hydrated electrons via an intermediate charge transfer state[13] or ionization of other solutes such as $K_4Fe(CN)_6$[14] or toluene.[15] However, none of these processes can be induced by visible light.

These points have led to a renewed search for efficient sources of hydrated electrons using as much of the solar spectrum as possible.[16–19] A recent approach, for example, has involved the photoexcitation of the localized surface plasmon resonance of aluminium and silver nanostructures.[5,18] The hot electrons produced following dephasing of the plasmon can be injected into the surrounding organic solvent thus producing solvated electrons. However, the many competing processes such as electron-electron scattering within the nanoparticles (NPs), which thermalize the electrons before transfer can take place mean that this process is destined to have a low efficiency. Another promising material for the production of hydrated electrons is hydrogenated diamond surfaces.[10,20–22] It has long been known that these surfaces have the intriguing property of effective negative electron affinity which means that an electron excited into the conduction band of the diamond can be directly emitted into vacuum[23]. This is mainly due to oriented dipoles at the H-terminated surface which guarantee that electrons that approach the surface are emitted. In the case that the material is immersed in water the electrons are emitted into solution forming hydrated electrons.[22,24] However, diamond is a very wide band gap material ($E_g$ = 5.5 eV) and thus only VUV photons ($\lambda$< 225 nm) can be used to directly excite electrons into its conduction band. As a matter of fact, it has been shown that exciting thin films of hydrogenated diamond with 213 nm light[22] and both hydrogenated and hydroxylated nanodiamonds with 225 nm light[24] can produce hydrated electrons. These hydrated electrons have been demonstrated to be utilizable to promote photochemical redox reactions.[8,25–27]

NPs of diamond can be synthesized by several processes involving shock compression, explosive detonation, laser ablation of non-diamond carbon and ball milling of larger diamond particles[28–30] and are now widely commercially available as powder or dispersion with particles size ranging between 20 and 3 nm depending of the synthesis procedures. Nanodiamond particles obtained by detonation are usually called Detonation Nanodiamonds (DNDs) and are composed of particles size ranging between 3 – 5 nm in diameter; DNDs have interesting properties such as extremely large surface-to-volume ratios,[31] biocompatibility,[32] and possible tailoring of the surface chemistry.[20] The dynamics of the production of hydrated electrons from sub-band gap excitation of suspensions of these NPs have been investigated both on the picosecond[24] and microsecond[14] timescales.

In a recent study, silver NPs in a diamond film on a silicon substrate were shown to exhibit $N_2$ reduction capabilities when illuminated with sub bandgap light.[33] Furthermore, other material formed by coating CVD diamond onto metal substrates of niobium[34,35] and tungsten[36] have shown promise for hydrated electron production and electron emission, respectively.

These works have encouraged us to combine plasmonic NPs with DND dispersed in water to enhance the generation of hydrated electrons due to interaction between the plasmonic resonance of the metal NPs and the nano diamonds. The aim is to exploit AuNPs as efficient visible light absorbers which, in turn, enhances the efficiency of the electron emission into water from the DND nanoparticles as described below. In this case, gold NPs were chosen to be coupled to the DND (AuNP@DNDs) in aqueous solution[37] and the production of hydrated electrons was monitored by a femtosecond all-optical pump-probe experiment in which the pump is used to excite the plasmon resonance and a white light probe is used to verify the formation of the hydrated electron by monitoring its optical absorbance centered at 720 nm.

## 2. EXPERIMENTAL SECTION

### 2.1 Synthesis

AuNP@DND were produced by a procedure developed in our lab and described in detail in refs.[38,39] In brief, the detonation nanodiamonds (DND) were purchased from the International Nanotechnology Center (ITC, USA) and the synthesis of the AuNPs takes place through the direct reduction of the gold ion complex on the surface of the nanodiamond which acts as a reducing agent. In order to increase the reducing power of the DND, it was previously subjected to a 12-hour reaction with an aqueous solution of $NaBH_4$ (with a weight ratio ND:NaBH4 of 1:2), after that the dispersion was washed until neutral pH and finally an equal volume of Au complex $10^{-4}$ M was added. The dispersion was sonicated for 30 min and then the reaction was stopped by separating the powdering product from the supernatant. After several cycles of water washing the sample was analyzed and structurally characterized. A more detailed account of the characterization of the AuNP@DND system investigated here can be found in ref.[40]

### 2.2 Scanning Electron Microscopy

A few microliters of AuNP@DND dispersion were spin-coated onto a Silicon <100> substrate. The dried samples of AuNP@DND were analyzed on a Field Emission Scanning Electron Microscope (FE-SEM) ZEISS SIGMA 300 (see Figure S1 and the discussion of the morphological analysis below). The morphological characterization has been carried out at an accelerating voltage of 10 kV and using the in-lens electron (SE) detector.

### 2.3 Dispersion particles-size distribution

Particle-size distribution of the DND and AuNP@DND water dispersion were measured using a Malvern Zetasizer Nano (Malvern Panalytical) and directly analyzed by Zetasizer software (Malvern Panalytical). The viscosity and dielectric constant of the samples were assumed to be equal to that for water at 25 C. Figure S2 reports the aggregate particles size distribution for both DND and AuNP@DND.

## 2.4 Femtosecond Transient Absorption Spectroscopy

The ultrafast transient absorption measurements were performed using a laser system consisting of a chirped pulse amplifier seeded by a Ti:Sa oscillator. A schematic of the optical set-up can be found in the supporting information (see Figure S3) together with details of the pump-probe experiment. In brief, the wavelength tunable pump pulses (350 – 600 nm) were from a commercial optical parametric amplifier (OPA) while the white light probe was generated by focusing the output of a homebuilt 1300 nm OPA into a sapphire crystal. The latter produced white light in the range from 500 to 1100 nm which was split into two equal parts one of which passes through the sample while the other is used as a reference to account for pulse-to-pulse fluctuations in the white light generation. The pump pulse is loosely focused (circular spot of diameter of 180 μm) onto the sample with an energy density from 1.0 mJ/cm$^2$ to 6.2 mJ/cm$^2$. The spot diameter of the probe pulse is much smaller (approx. 100 μm) and its delay time with respect to the pump pulse is scanned in time by varying the length of its optical path. The instrument response function was measured to be approximately 50 fs. All measurements were performed in air at room temperature. Further details of the set-up can be found in previous publications.[41,42] It should be noted that during the data acquisition both the AuNP@DND and the bare DND suspensions undergo partial precipitation with a reduction of the signal by <5% during an acquisition. Simple agitation of the suspension returns the signal level to the original level and using this procedure the signal level remained constant for multiple days of illumination.

## 3. RESULTS AND DISCUSSION

### 3.1 Calibration of Hydrated Electron Detection Efficiency

To verify the optical response of our pump-probe set-up to hydrated electrons we first measured the production of the HEs in a 0.2 M aqueous solution of KI. Photoexcitation of this solution with 350 nm (3.54 eV) permits the ionization of the I⁻ in solution with the subsequent formation of the hydrated electron through initial population of a charge transfer to solvent (CTTS) state.[13] The transient absorption map is shown in Figure 1 a). As expected, the CTTS state has a slightly different and broader absorption profile which evolves into the hydrated electron absorption which is spectrally narrower and peaked at 720 nm (see Figure 1 c)). Indeed, by subtracting the temporal evolution of the photoinduced absorption (PIA) at 570 nm (where the CTTS state absorption dominates) from the temporal evolution of the PIA at 720 nm (where both hydrated electrons and CTTS PIAs combine) it is possible to isolate the dynamics of hydrated electron generation. In Figure 1 b) one can see that the signal related to hydrated electrons appears with a delay of about 2 ps and reaches a maximum at 4.5 ps.

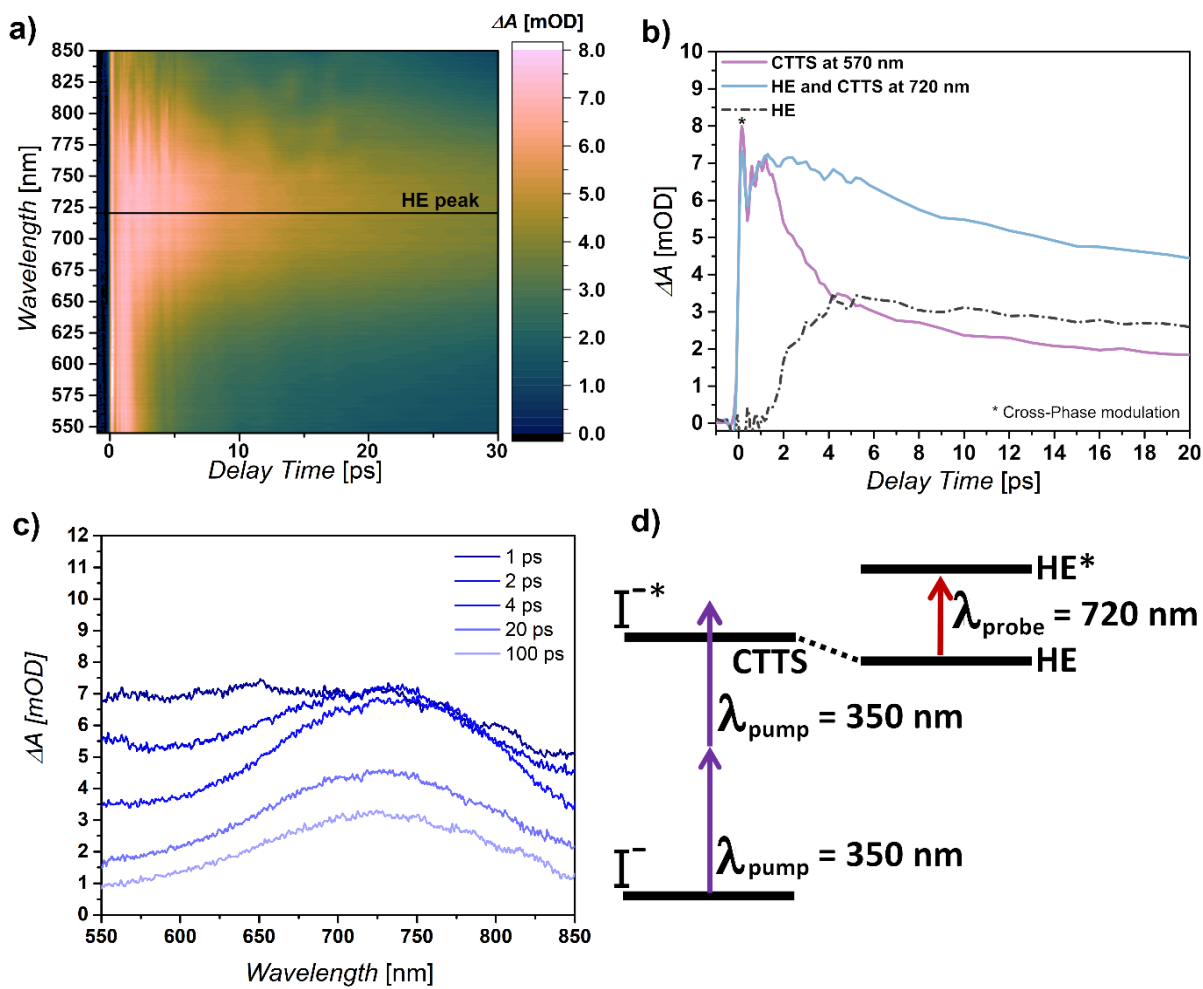

**Figure 1.** a) Transient absorption map of a 0.2 M KI solution in water with pump radiation at 350 nm and a fluence of 5.5 mJ/cm$^2$, b) Temporal cuts of the PIA at 570 nm (CTTS state) and 720 nm (CTTS and hydrated electron) and the difference between the two (hydrated electrons alone), c) Spectral cuts showing the evolution of the PIA in the first 100 ps after excitation and d) scheme indicating the two-photon excitation and one-photon probe of the PIA.

The power dependence of the signal shows that the process is not linear and behaves as a P$^n$ process where n = 1.49 (see Figure S4). This is consistent with the fact that the lowest lying CTTS state is 5.5 eV above the ground state and therefore not accessible with a single photon of wavelength 350 nm (Figure 1d)). A two-photon excitation process usually exhibits a power dependence with an exponent of n = 2, while in this case the reduced value of n suggests a partial saturation of the excitation process.

## 3.2 Morphological characterization of the self-assembled gold nanoparticles and nanodiamonds

The self-assembled gold nanoparticles and nanodiamonds (AuNP@DND) which form the material for this study were synthesized as described in the methods. For the SEM characterization a few microliters of the AuNP@DND dispersion were spin coated onto a Si <100> substrate and the resulting SEM images are reported in Figure 2a) and Figure S1. From this morphological analysis it can be seen that the DNDs greatly outnumber the AuNPs with the formation of clusters of DNDs with a spongy like structure (see Figure S1) decorated by quasi-spherical Au NPs.

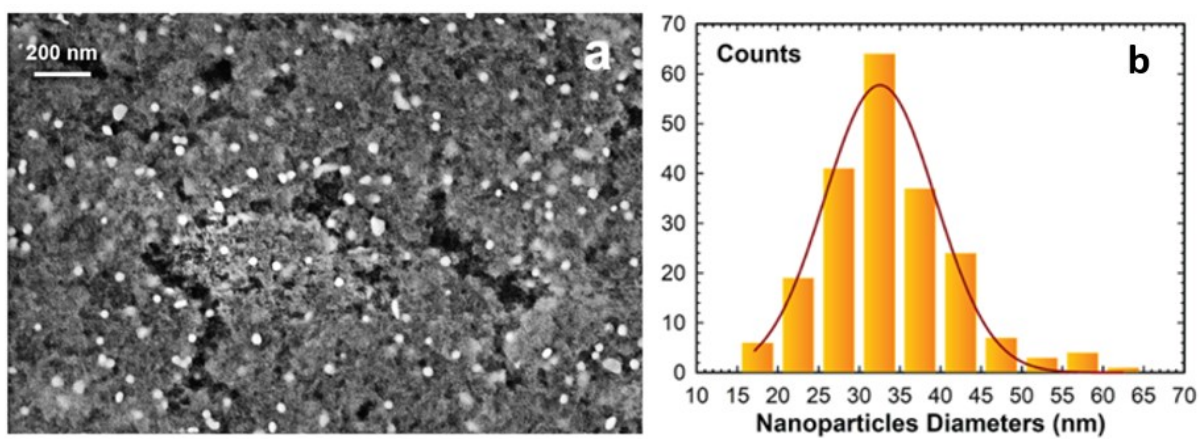

**Figure 2.** Morphological study of self-assembled gold nanoparticles and nanodiamonds (AuNP@DND). a) SEM image illustrating the surface morphology. b) Histogram showing the frequency distribution of the AuNP size.

The SEM image (Figure 2 a)) shows bright nanometric spots overlapping a flat surface indicating the presence of well separated AuNPs of slightly different dimensions (see Figure 2 b)). Quantitative morphometric measurements and a possible topological relationship among the NPs have been investigated. A 2D contour map of 206 AuNP distribution has been quantified on a probed area of 2285 nm by 1524 nm. The morphometric analysis has identified NPs of quasi-spherical shape. The measured diameters of the NPs provided an average value of 33 ± 15 nm (Figure 2 b)).

**3.3 Emission of electrons into water following visible light irradiation of AuNP@DND**

Figure 3 a) shows the transient absorption map of AuNP@DND in aqueous dispersion where the concentration of Au is estimated to be 80 ± 20 µg/ml from absorbance measurements. One can again observe the appearance of a PIA band centered at about 725 nm following the photoexcitation of the plasmon resonance of the AuNPs (see stationary absorbance of the solution in Figure 5). It should be noted that no PIA at 725 nm was observed when exciting pure water or DND dispersed in water (concentration 0.5 mg/ml) with the same fluence of 530 nm light (see the transient absorption map of DND in water in Figure S5 of the supporting information.). This shows that the localized surface plasmon resonance of the AuNPs is involved in the emission of electrons. This point will be discussed in detail below.

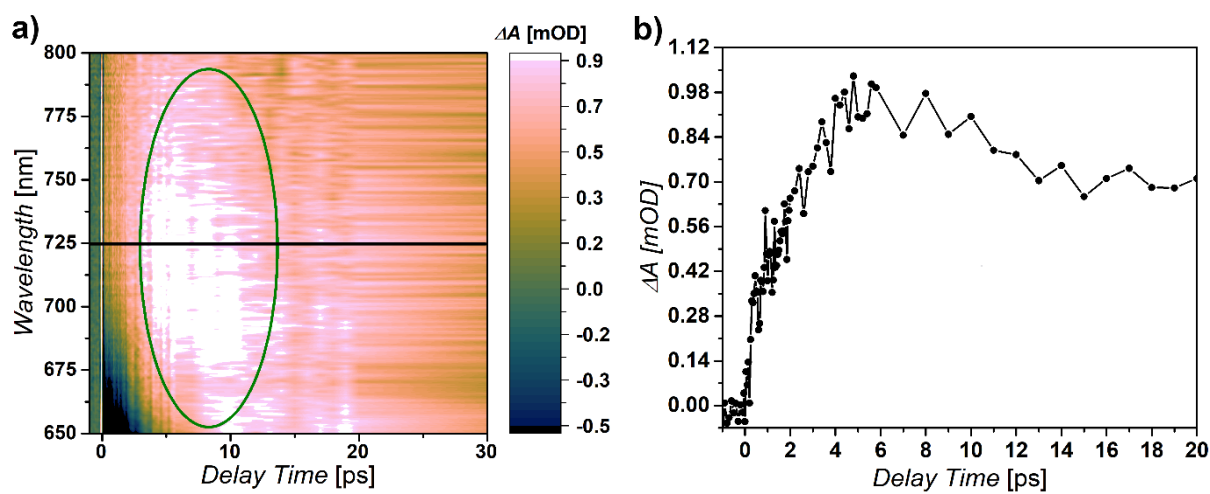

**Figure 3.** a) Transient absorption map of AuNP@DNDs following excitation at 500 nm and a fluence of 6.2 mJ/cm$^2$. The white area represents the signal of the hydrated electrons (circled in green). It has been saturated to emphasize the feature. b) Temporal behavior of the PIA signal related to the hydrated electrons, averaged between 700 – 750 nm.

Since in the spectral region between 700 nm and 750 nm we can neglect any other transient signals and the hydrated electron is fully formed after 5 ps, we can assume that the transient signal at these wavelengths and at this delay is only due to the induced absorbance of the hydrated electrons. To estimate the concentration of hydrated electrons $n$, we used the Lambert-Beer formula:

$$n = \frac{A(OD) \cdot N_A}{\epsilon(cm^2 \cdot mol^{-1}) l(cm)}$$

where $l$ is the cuvette length, $N_A$ is the Avogadro constant and A(OD) and $\epsilon$ are respectively the absorbance and the molar absorption coefficient of the hydrated electrons ($\epsilon = 1.90 \cdot 10^7$ mol$^{-1}$ cm$^2$)[2]. Table 1 reports the hydrated electron density produced in the KI solution and in the AuNP@DND sample. The injection efficiency ξ is then evaluated considering the ratio between $n$ and the number of photons per unit volume ($N_{ph}$).

| Sample | ΔA [OD] | λp [nm] | Fluence [mJ/cm$^2$] | n [$10^{15}$ cm$^{-3}$] | $N_{ph}$ [$10^{17}$ cm$^{-3}$] | ξ [%] |
|---|---|---|---|---|---|---|
| KI | 0.007 | 350 | 5.5 | 2.22 | 0.97 | 2.29 |
| AuNP@DnD | 0.00098 | 500 | 6.2 | 0.31 | 1.55 | 0.20 |

**Table 1:** Values for the efficiency of hydrated electron formation extracted from the transient absorption data. λp is the wavelength of the pump radiation.

To verify that the signal in the 700 – 750 nm range is indeed due to a PIA caused by the formation of hydrated electrons we have added an electron scavenger to the AuNP@DND solution to capture the hydrated electrons and thus quench the electron PIA signal. To do so we have diluted the same quantity of sample (50 µl) with either 50 µl of pure water or 50 µl of a 1 M solution of NaNO$_3$. The nitrate ions are very efficient electron scavengers even on femtosecond timescales.[43] The result is shown in Figure 4 where it can be seen that the NO$_3^-$ ions quench 40-50% of the hydrated electrons, which is consistent with the values achieved in photo-induced radiolysis of water.[43] This confirms the assignment of the signal in the 700-750 nm range to the hydrated electrons as the addition of NO$_3^-$ to the solution has no significant effect on the plasmon signal. It should be noted that additional signal in the 650 -700 nm range may be due to incomplete hydration of the injected electron which could distort the shape of the PIA, or to the simple overlap of the two contributions from plasmon and solvated electrons. In both cases, the effect of NaNO3 is the reduction of the overall signal in both ranges.

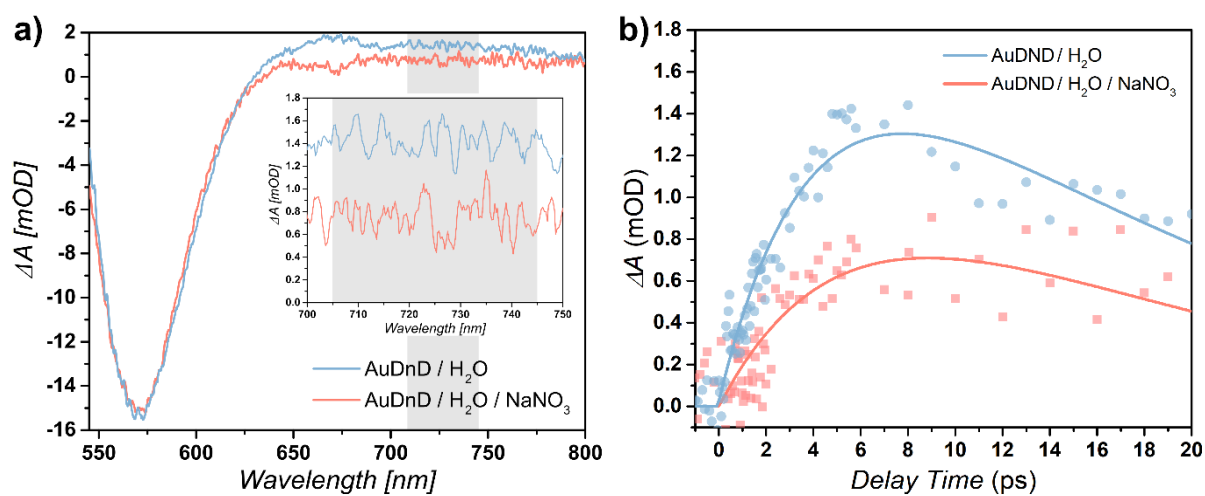

**Figure 4.** a) Effect of the addition of an electron scavenger (NaNO$_3$) to the AuNP@DND sample on the formation of the hydrated electron as shown by the difference between the transient absorption of the AuNP@DND/H$_2$O sample and that of the AuNP@DND/(H$_2$O+NaNO$_3$) sample. b) Temporal behavior of the hydrated electron PIA averaged over the shaded region in a) in the AuNP@DnD/H$_2$O unquenched sample (blue) and in the AuNP@DND/(H$_2$O+NaNO$_3$) quenched sample (red). The full data of the transient absorption from which these data were extracted are shown in Figure S7.

A further control experiment to determine the role of the DND in the electron emission process was to excite a suspension of the citrate stabilized 10 nm AuNPs in a 1 mm cell by a pulse of 550 nm light and to monitor the change of the transmission of the sample as a function of the time delay up to 25 ps between the pump and white light probe (350 – 780 nm) to search for solvated electrons emitted directly from the AuNPs. The transient absorption map and its analysis are discussed in detail in the supporting information (see section S1.6 and Figure S6), with the main conclusion being that solvated electrons could not be observed following plasmonic excitation of the bare AuNPs within the signal to noise of our set-up. This is consistent with the conclusion of Al-Zubeidi *et al.*[18] that emission from gold nanostructures following photoexcitation on the plasmon resonance is energetically forbidden.

To further investigate the role of the plasmon resonance in the formation of the hydrated electrons we have pumped the AuNP@DND sample at a range of wavelengths across the plasmon resonance of the AuNPs and used the resulting transient absorption to extract the wavelength dependent injection efficiency of hydrated electrons using the method outlined above. In the case that the efficiencies were to be equal across the plasmon resonance this graph should show a flat line for the relative efficiency. Instead, as shown in Figure 5, this ratio increases strongly to the higher energy side of the plasmon resonance while on the low energy side of the plasmon resonance (> 600 nm) no hydrated electron signal was observed because the signal intensity goes below the sensitivity of the experiment.

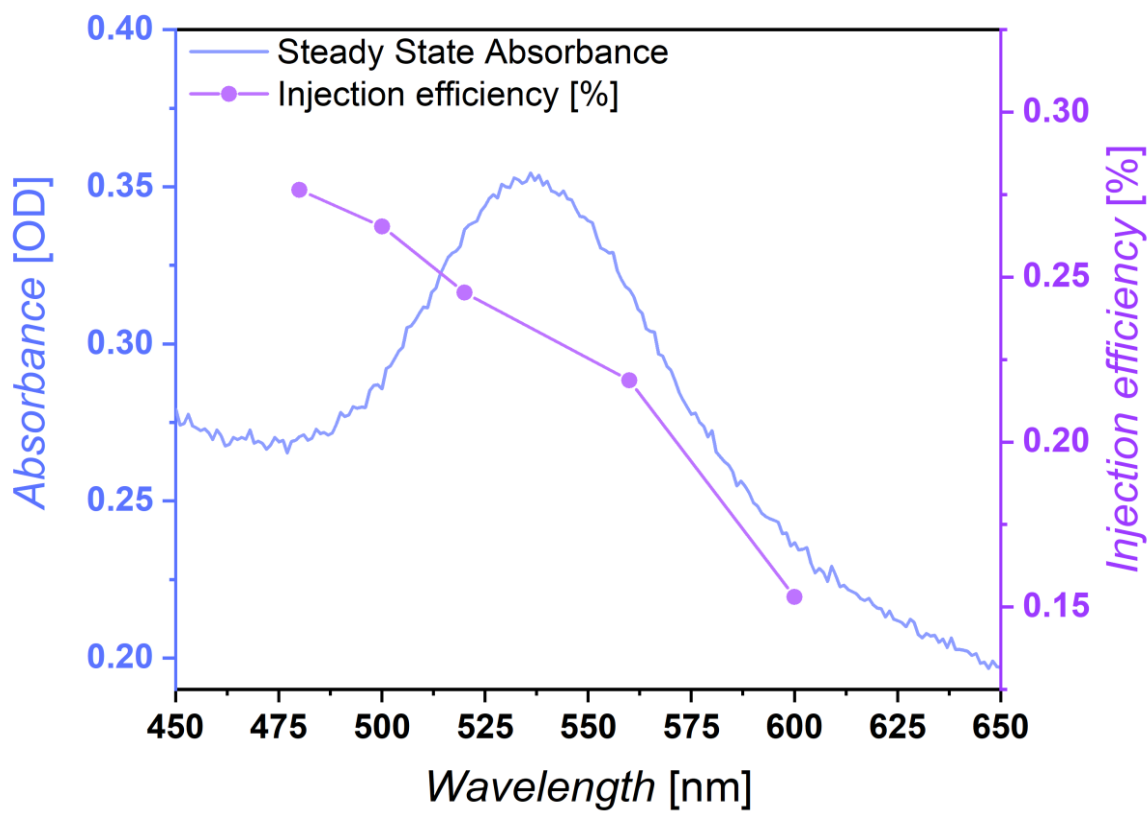

**Figure 5.** Stationary absorbance of the AuNP@DND sample (left axis in blue) measured in a 1 mm cell and estimated hydrated electron injection efficiency (right axis in violet) at different pump energies (475 nm, 500 nm, 525 nm, 550 nm, 600 nm) and considering the ratio between the density of the hydrated electrons at a delay time of 5 ps (maximum of the hydrated electron PIA) and $N_{ph}$, the number of impinging pump photons per unit volume.

So, while the plasmon does modulate the electron injection efficiency, it is not the only factor involved. To gain some insight into the underlying mechanism we consider a previous work on the generation of hydrated electrons from dispersed DND aqueous solutions by irradiating with 225 and 400 nm light.[24] In that work the authors postulated that the sub bandgap injection of electrons from DND into water was due to the presence of sp$^2$ islands (which they called fullerene-like reconstructions) on the surface of the NPs. These islands result in the presence of both filled and unfilled defect-like states between the VB and the CB of the diamond[24] which can give rise to a weak absorption in the visible region of the spectrum.[44] They suggest that this absorption is involved in the injection of the electrons from the DND into the water although it is not clear whether this is a transition between $sp^2$ states followed by the formation of a charge transfer state (as was observed in the case of I$^-$ described in section 2.1) or possibly a direct injection of the $sp^2$ electrons into the water layer. The above authors preferred a molecular like description of the process and performed DFT calculations on small diamond like clusters with different quantities of $sp^2$ carbon on the surface to confirm their hypothesis. Ristein et al.[45], on the other hand, who originally proposed this kind of mechanism to explain sub-bandgap photoelectron emission from H-terminated single crystal diamond, described the process in terms of inhomogeneous emission from semi-metal graphitic patches on the surface of the diamond. In their view, electrons from the $\pi$ band of the graphitic patches were emitted directly into the vacuum. A similar mechanism, although involving a multiphoton process, was also proposed for the emission of electrons from diamond-coated tungsten tips using femtosecond laser pulses.[36] In the present case, a power dependence study of the HE signal revealed a linear dependence of the PIA signal on the power of the 500 nm pump (see Figure S8). This indicates that the injection process observed here involves a single photon.

In any case, the physical system investigated here is very complex and a clear picture of the process is difficult to define. Indeed, when aiming to explain the experimental evidence we have observed, it is important to note that visible light absorption and electron emission into water already occur on the bare DND nanoparticles. While our experiments were not sensitive enough to detect hydrated electrons after irradiation with 530 nm light of an aqueous suspension of DNDs, other works have observed the formation of hydrated electrons after irradiation with 254 nm and 400 nm[24] and visible light absorption of bare DNDs have been observed in the 400 – 700 nm range.[44] The mechanism for electron emission following sub bandgap photoexcitation is linked to the tail of absorption into the visible region observed in the steady state absorption of bare DNDs (see Figure S9 in the supporting information). So while the process likely occurs at 530 nm it is probably so inefficient that we cannot observe it. In our set-up we require the presence of both the AuNPs and the DND in order to observe the hydrated electrons as neither solution alone pumped in the visible region produces sufficient HEs to be observed in transient absorption.

Based on the above observation we suggest that the addition of AuNPs simply enhances the efficiency of these already existing processes in a SERS-like manner.[46] The simplest mechanism by which the AuNPs can enhance a process in a resonant manner (cases in which there is a spectral overlap between the SPR excitation and the transition to be enhanced) is by enhancement of the local electric field of the light on the surface of a NP due to oscillations of the conduction electrons. Absorption at the surface of the NP can be greatly enhanced by this mechanism with enhancement factors in excess of 1000 being observed.[47] More complex mechanisms, such as hot electron injection from the Au to the diamond,[48] chemical enhancement (in analogy to SERS chemical enhancement)[46] or resonant energy transfer[49] cannot be ruled out *a priori* but are not needed to explain the enhancement observed.

A final important observation to be explained is the value of the rise time of the hydrated electron signal which reaches its maximum at 5 ps after photoexcitation. This phenomenon is consistent with the solvation time of the hydrated electrons emitted from I$^-$ where we observe that when the contribution of the CTTS state is subtracted from the temporal profile the signal of the hydrated electron reaches a maximum after more than 4 ps. Such a timescale is consistent with the electron being emitted from the surface of the diamond into a CTTS state followed by relaxation into a hydrated structure.

Although in this work only short time dynamics of the hydrated electrons have been measured (< 20 ps), previous studies have shown that hydrated electrons emitted from DNDs following

UV irradiation can live up to tens of microseconds, a sufficient time to perform a reduction reaction and degrade a range of dangerous organic substances[7] such as perfluorooctanesulfonate (PFOS) which are persistent pollutants found globally in water supplies.[27] We expect that the AuNP@DND nanocomposites presented here can extend such activity into the visible light range. Furthermore, both AuNPs and DNDs are biocompatible and therefore, in addition to the industrial applications of $CO_2$ and $N_2$ reduction and wastewater treatment, applications involving the production of hydrated electrons in biological environments can be explored. Finally, the efficiency of the process can be optimized by tuning the AuNP to DND ratio of the suspensions, as well as the termination and surface chemistry of the DNDs (e.g., by performing a specific process of hydrogenation).

## 4. CONCLUSIONS

We have successfully increased the efficiency of visible light induced production of hydrated electrons from detonated nanodiamonds (DNDs) dispersed in water by growing gold nanoparticles onto which the DNDs attach (AuNP@DND). The production of hydrated electrons was monitored by the observation of a transient absorption signal at 700-750 nm which could be efficiently quenched by the addition of an electron scavenger ($NO_3^-$) to the aqueous dispersion of AuNP@DND. We suggest that our observation is made possible by the excitation of states involving $sp^2$ hybridized islands on the DND surface which are strongly increased by the presence of the AuNPs most likely through local electric field enhancement on the surface of the NPs. This composite material has the potential to have a major impact on the homogeneous catalysis of energy intensive reactions such as $CO_2$ reduction or even $N_2$ fixing as it could provide an efficient and non-toxic source of the ideal reducing agent that is the hydrated electron from solar excitation. While a relatively low efficiency is achieved in this first attempt to use AuNP@DND composites, we expect that the understanding of the electron injection process provided here together with modifications of the materials involved will lead to a significant increase in the injection efficiency achievable.

ASSOCIATED CONTENT

**Acknowledgments**

R.M. acknowledges financial support from the NationalQuantum Science Technology Institute within PNRR MURproject PE0000023-NQSTI. S.O. acknowledges the GrantMUR Dipartimento di Eccellenza 2023-27 X-CHEM project"eXpanding CHEMistry: implementing excellence in researchand teaching". P.O.K. acknowledges an Italian Ministerial grantPRIN PNRR 2022 Grant No. P2022ZHCT3 financed by theEuropean Union - Next Generation EU.